\documentclass[sigconf]{acmart}
\pdfoutput=1
\usepackage{soul}
\usepackage{array}
\usepackage{color,soul}
\usepackage{enumitem}   
\usepackage{stfloats}

\AtBeginDocument{%
  }

\setcopyright{none} 

\acmDOI{How.to.remove.this?(Not.ACM)}

\acmConference[`6th Championship Branch Prediction (CBP)']{6th Championship Branch Prediction}{June 21, 2025}{Tokyo, Japan}

\acmISBN{}                    
\acmDOI{}                     
\acmYear{}                  
\begin{document}

\title{Taming Wild Branches: Overcoming Hard‑to‑Predict Branches using the Bullseye Predictor}

\author{Emet Behrendt}
\email{emet@student.ubc.ca}
\affiliation{
  \institution{The University of British Columbia}
  \city{}
  \country{Canada}
}

\author{Shing Wai Pun}
\email{swpun@student.ubc.ca}
\affiliation{
  \institution{The University of British Columbia}
  \city{}
  \country{Canada}
}

\author{Prashant J. Nair}
\email{prashantnair@ece.ubc.ca}
\affiliation{
  \institution{The University of British Columbia}
  \city{}
  \country{Canada}
}


\begin{abstract}
Branch prediction is key to the performance of out‑of‑order processors. While the CBP-2016 winner TAGE-SC-L combines geometric-history tables, a statistical corrector, and a loop predictor, over half of its remaining mispredictions stem from a small set of hard-to-predict (H2P) branches. These branches occur under diverse global histories, causing repeated thrashing in TAGE and eviction before usefulness counters can mature. Prior work shows that simply enlarging the tables offers only marginal improvement.

We augment a 159 KB TAGE-SC-L predictor with a 28 KB H2P-targeted subsystem called the \emph{Bullseye} predictor. It identifies problematic PCs using a set-associative H2P Identification Table (HIT) and steers them to one of two branch-specific perceptrons, one indexed by hashed local history and the other by folded global history. A short trial phase tracks head-to-head accuracy in an H2P cache. A branch becomes perceptron-resident only if the perceptron's sustained accuracy and output magnitude exceed dynamic thresholds, after which TAGE updates for that PC are suppressed to reduce pollution. The HIT, cache, and perceptron operate fully in parallel with TAGE-SC-L, providing higher fidelity on the H2P tail. This achieves an average MPKI of 3.4045 and CycWpPKI of 145.09.
\end{abstract}


\maketitle

\section{Introduction}

Modern out-of-order CPUs depend heavily on aggressive speculative execution, where each fetched instruction block is predicted to either fall through or take a branch long before the branch condition is resolved. A single misprediction triggers a pipeline flush, drains the front-end, and requires re-fetching along the correct path, typically incurring a penalty of 15–30 cycles. As a result, the design and accuracy of the branch predictor have direct and significant performance implications, affecting instruction throughput, energy efficiency, and speculation-dependent microarchitectural features.

Branch prediction accuracy remains a first-order design constraint, shaping front-end bandwidth, retirement rates, and security mitigation strategies. Over the past three decades, predictors have evolved from simple bimodal schemes to sophisticated hybrid, perceptron-based, and multi-component designs. The TAGE-SC-L predictor, the CBP-2016 competition winner, combines tagged geometric history tables (TAGE) with a statistical corrector and a loop predictor to cover a broad spectrum of control-flow behaviors. However, despite this architectural complexity, TAGE-SC-L still leaves substantial accuracy on the table, especially for a small but critical subset of dynamic branches.

Despite TAGE-SC-L's advanced use of long-range geometric histories, statistical correction, and loop-specific components, its residual mispredictions remain highly skewed. Lin et al.~\cite{Lin_2019} showed that in a 30-million-instruction SPEC-INT 2017 trace, over 50\% of mispredictions stem from fewer than ten static hard-to-predict (H2P) branches. These branches exhibit volatile control-flow behavior, with dynamic instances appearing under widely varying global histories. As a result, TAGE’s tag-matching struggles to retrieve consistent predictor state, leading to frequent allocations with low confidence. Since usefulness counters only increment on correct predictions, these entries are quickly evicted, creating a self-reinforcing cycle of misprediction, reallocation, and eviction that prevents the predictor from learning stable correlations.

An H2P branch may see hundreds of distinct TAGE entries within a single millisecond of execution, none surviving long enough to gather confidence. Intuitively, one might expect that adding more storage, such as enlarging each component table or appending extra history lengths, would amortize the thrashing. Yet idealized studies by Seznec~\cite{Seznec_Idealistic_GTL} and the independent branch‑runahead framework of Pruett et al.~\cite{pruett_2021_branch_runahead}, demonstrate that even a hypothetical TAGE‑SC‑L with unbounded capacity converges only marginally beyond the accuracy of the published 64 kB design. Capacity alone cannot compensate for the entropy inherent in the branch's context. Thus, the predictor tends to lack a representation capable of generalizing across diverging histories that precede each H2P instance.

We introduce Bullseye, a compact H2P subsystem that augments a 159.3 kB TAGE‑SC‑L to address these challenges. Bullseye first pinpoints the few branches that dominate the residual error and then hands them off to branch‑specific perceptrons. Detection is handled by the H2P Identification Table (HIT), a set‑associative array that records each static branch’s execution and misprediction counts. A branch becomes H2P‑active only after it surpasses adaptive thresholds on executions, mispredictions, and accuracy limits that tighten as the number of active H2P branches grows. Real workloads seldom activate more than eight to ten PCs simultaneously, so the HIT remains small and latency‑neutral.

Once Bullseye flags a program counter, the branch enters a brief trial phase: TAGE‑SC‑L and two perceptron engines predict in parallel while an H2P cache tracks their head‑to‑head accuracy. Suppose either perceptron maintains higher accuracy than TAGE‑SC‑L and produces outputs above a dynamic magnitude threshold. In that case, Bullseye promotes the branch to perceptron‑resident status and suppresses further TAGE updates for that PC, eliminating table pollution. Branches that fail the trial fall back to the HIT with negligible overhead.

Bullseye's prediction engines comprise two lightweight perceptrons that capture patterns that TAGE's geometric indexing misses. One leverages hashed‑window local history for fine‑grained, per‑branch behavior; the other uses a folded global history vector to learn long‑range correlations. During each fetch, the HIT probe, both perceptron evaluations, and the baseline TAGE‑SC‑L lookup proceed in parallel. A single‑cycle arbiter selects a perceptron's output when its confidence dominates; otherwise, TAGE‑SC‑L supplies the prediction. By confining extra complexity to the stubborn H2P tail, Bullseye sharpens overall accuracy while preserving the critical‑path speed of the underlying predictor.

\section{Background and Related Work}

Modern branch predictors can be broadly classified into TAGE-based and perceptron-based categories.

\subsection{The TAGE-based Predictor}
The TAGE predictor architecture combines a tagless bimodal table of PC-indexed 2-bit counters with multiple tagged components indexed using geometrically increasing history lengths. Each tagged entry includes a prediction counter, a partial tag, and a usefulness counter. The final prediction is selected from the tagged component with the longest matching history. TAGE-SC-L enhances this core design with two additional components: (i) a statistical corrector (SC) that aggregates predictions from biased, global, and local history tables to override low-confidence TAGE outcomes, and (ii) a loop predictor (L) optimized for detecting constant-iteration loops~\cite{Seznec_TAGE}. TAGE-SC-L was the winning submission in CBP-2016 and serves as the baseline for the predictor proposed in this work.

\subsection{The Perceptron-Based Predictor}
Perceptron-based predictors take a fundamentally different approach by framing branch prediction as a form of single-layer neural inference. Each prediction is computed as a weighted sum of recent branch outcomes combined with a bias term; the sign of the resulting sum determines whether the branch is predicted as taken or not taken~\cite{Perceptrons_Jimenez}. Weights are updated at retirement based on the correctness of the prediction, allowing the perceptron to learn long-term correlations that traditional counter-based predictors typically miss. While perceptrons are well-suited for linearly separable patterns, they introduce higher latency and storage costs, especially as history lengths grow and weight vectors expand.

\subsection{Why use H2P-Tailored Predictors?}
While continued process scaling has enabled significantly larger branch predictor budgets, idealized studies reveal diminishing returns in accuracy per kilobyte as predictor size increases~\cite{Seznec_Idealistic_GTL}. Even with sophisticated designs like TAGE‑SC‑L, detailed execution traces show that a small number of hard-to-predict (H2P) branches account for a disproportionate share of mispredictions, limiting overall IPC gains~\cite{Lin_2019}. These insights motivate hybrid predictor architectures, such as the one proposed in this work, that retain TAGE-SC-L's low-latency, high-throughput backbone while integrating lightweight perceptron models explicitly targeted at these high-impact, difficult-to-model branches.

\section{Bullseye: High-Level Design Overview}

\begin{figure}[htbp]
    \centering
    \includegraphics[width=0.8\linewidth]{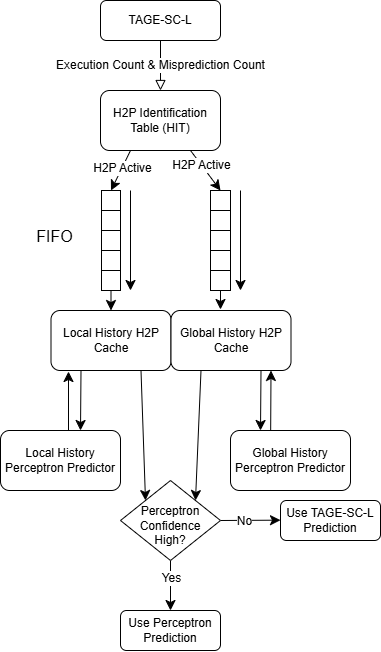}
    \caption{A high-level overview of the architecture of the Bullseye prediction system.}
    \Description{A high-level overview of the architecture of the Bullseye prediction system.}
    \label{fig:h2p_predictor}
\end{figure}

Figure~\ref{fig:h2p_predictor} shows the control flow through Bullseye's H2P pipeline. Execution begins on the baseline TAGE‑SC‑L, which supplies both predictions and two running statistics to the H2P Identification Table (HIT). Namely, the per‑PC execution count and misprediction count. When a branch's counters exceed Bullseye's adaptive thresholds, the HIT flags it H2P‑active and enqueues the PC, in FIFO order, into two small tag‑RAMs: the local‑history H2P cache and the global‑history H2P cache. These caches hold only a handful of H2P PCs observed in practice ($\le$10), ensuring constant‑time look‑ups without inflating front‑end latency.

On every fetch, both caches probe in parallel. A hit launches the corresponding local‑history or global‑history perceptron predictor, which reads its weights, produces an output, and forwards the result to a confidence arbiter. If either perceptron’s magnitude and running win rate exceed Bullseye’s dynamic threshold, the arbiter overrides TAGE‑SC‑L with the perceptron prediction; otherwise, the system returns to the baseline forecast. This selective substitution confines the extra latency of neural evaluation to the rare H2P branches while preserving the single‑cycle critical path for the typical case, thereby sharpening overall accuracy without compromising pipeline depth.

\section{Bullseye Predictor Operation}

Our branch predictor uses a naively scaled-up 159.3 kB version of the provided TAGE-SC-L with additional logic to identify and predict H2P branches. This section details the functionality of each key component in the branch predictor.

\subsection{Hard-to-Predict Identification Table (HIT)}
\label{sec:hit}

Bullseye locates key H2P branches with a small, set‑associative
\emph{Hard‑to‑Predict Identification Table (HIT)}. Each HIT entry maintains three running statistics for a static branch~\(b\):

\begin{itemize}[leftmargin=*]
  \item \(\text{Exec}(b)\): cumulative dynamic executions,
  \item \(\text{Mispred}(b)\): cumulative TAGE‑SC‑L mispredictions,
  \item \(\text{Acc}(b)=1-\text{Mispred}(b)/\text{Exec}(b)\): running accuracy.
\end{itemize}

\subsubsection{Running Statistics}
Let \(N_{\text{H2P}}\) denote the \emph{current} number of branches already
classified as hard‑to‑predict (H2P‑active) and therefore resident in the
perceptron layer. A new branch becomes H2P‑active, and is queued into both the local‑ and
global‑history H2P caches, when it first satisfies the adaptive rule set in Equations~\eqref{eq:h2p_def}\,(a–d):
\begin{subequations}\label{eq:h2p_def}
\begin{align}
\text{Exec}(b)      &\;\ge\; 2048 + 16\,N_{\text{H2P}},  \tag{\ref{eq:h2p_def}a}\label{eq:h2p_exec}\\
\text{Mispred}(b)   &\;\ge\; 256,                        \tag{\ref{eq:h2p_def}b}\label{eq:h2p_misp}\\
\text{Acc}(b)       &\;<\;  f\!\bigl(N_{\text{H2P}}\bigr).   \tag{\ref{eq:h2p_def}c}\label{eq:h2p_acc}
\end{align}
\begin{equation}
f(N)=
\begin{cases}
1 - 0.01\,N/32,          & N < 32, \\[2pt]
0.95 - 0.01(N-32),       & 32 \le N \le 71, \\[2pt]
0.60,                    & N > 71 .
\end{cases}
\tag{\ref{eq:h2p_def}d}\label{eq:h2p_f}
\end{equation}
\end{subequations}

\subsubsection{Leveraging the Statistics}
Equation~\eqref{eq:h2p_exec} raises the execution threshold by \(16\) for every additional H2P‑active branch, ensuring that transient bursts cannot flood the perceptron layer. Equation~\eqref{eq:h2p_misp} enforces a fixed lower bound of \(256\) mispredictions, filtering out seldom‑executed yet highly biased branches. The piece‑wise function \(f(N)\) in Equation~\eqref{eq:h2p_f} tightens the accuracy ceiling as \(N_{\text{H2P}}\) grows, dropping from \(>95\%\) when the perceptron
layer is empty to \(60\%\) at saturation (\(N_{\text{H2P}}>71\)). Because empirical workloads rarely exceed \(N_{\text{H2P}}\!\le\!10\), the HIT
remains compact.

\subsection{H2P Cache: Trial, Admission, and Eviction}
\label{sec:h2p_cache}

After the HIT flags a branch as \emph{H2P‑active}, its PC is inserted into both
the local‑ and global‑history perceptron engines and into a small, fully‑
associative \textbf{H2P cache}. The cache serves two purposes: (i) it records which PCs are currently predicted by Bullseye's perceptrons and (ii) it mediates a head‑to‑head “trial” between each perceptron and the baseline TAGE‑SC‑L.

\subsubsection{Trial phase}
    A newly admitted entry is granted a \emph{warm‑up window} of \(\mathbf{512}\) dynamic occurrences to train the perceptron weights. During this window, the branch is never considered for eviction. During and after the trial phase, two counters are used to determine H2P prediction confidence. Every prediction updates a saturating \textit{relative performance} counter based on the winning predictor. As the performance metric saturates, the stability of the \textit{relative performance} counter is then measured with a \textit{confidence counter} with \textit{linear growth and exponential decay}. Confidence is incremented by \(C\!\leftarrow\!\min(C{+}1,\,255)\) if the current \textit{relative performance} trend continues and  \(C\!\leftarrow\!C/2\) if the update goes against the running trend. This creates a policy that slowly rewards sustained predictor superiority yet quickly penalizes failing branches.

\subsubsection{Eviction policy}
A branch is evicted when either (i) the \textit{confidence counter} has saturated in favor of TAGE-SC-L or (ii) the entry is not referenced for \(2^{16}\) dynamic branches (``stale'' timeout). Eviction occurs only when at least one HIT‑qualifying branch is waiting to enter, thus guaranteeing high utility for every occupied slot.  Branches evicted from the cache revert to standard TAGE‑SC‑L prediction but may re‑enter if they later satisfy the H2P criteria in Equations~\eqref{eq:h2p_def}\,(a–d).

This gating mechanism ensures that Bullseye deploys its perceptron resources \emph{only} where they continue to beat the geometric core while bounding both storage and latency.  Empirically, no more than 10 PCs reside in the cache simultaneously, keeping look‑ups single‑cycle and energy‑efficient.

\subsection{Hashed‑Window Local‑History Perceptron}
\label{sec:local_perceptron}

Bullseye's first neural component targets correlations in branch's \emph{local history}. For each H2P‑active PC, the predictor constructs a feature vector whose \(i^{\text{th}}\) element is the parity of a fixed‑width \emph{window} \(W_i\) of the branch's outcome history, where window sizes grow with age (e.g., \(W_0{=}4\), \(W_1{=}8\), \(W_2{=}16\), \ldots). Successive windows start at offsets separated by a constant \textit{stride} \(S\), ensuring non‑overlapping coverage of up to a few
hundred past outcomes.

Each feature is mapped to two independent weight words by the hash \(h_i\bigl(\text{PC},W_i\bigr)\), implemented as a 32‑bit XOR‑shift scrambler. Dual hashing mitigates aliasing: a collision in one weight location is usually resolved by the second. The perceptron output is the integer sum of the selected weights plus a bias term. Finally, the sign of the output determines the prediction.

Weights are updated using the \emph{dynamic‑threshold} rule of Seznec and Vintan’s O‑GEHL predictor~\cite{Seznec_O_GEHL}. Specifically, a global threshold \(\theta\) tracks the absolute output magnitude at misprediction time and adapts toward the smallest value that keeps training activity near 50\%. A weight \(w\) is incremented (decremented) when the actual outcome is taken (not‑taken) \emph{and} \(|\text{output}| \le \theta\), enabling fast convergence without saturation. This hashed‑window design yields high resolution on recent local patterns while maintaining low storage per perceptron.

\subsection{Folded Global‑History Perceptron}
\label{sec:global_perceptron}

The global‑history perceptron captures long‑range, cross‑branch correlations. The feature vector consists of the most recent \(H_g\) outcomes from the \emph{global} branch history register, folded by XOR into a fixed \(W_g\)-bit index; each bit selects a single signed weight word. With a minimal number of weights, the global history perceptron size is an order-of-magnitude smaller than the local model.

Prediction and learning follow the same \emph{dynamic‑threshold} rule outlined in Section~\ref{sec:local_perceptron}. Specifically, the perceptron sums its \(W_g\) signed weights and compares the magnitude to a shared, runtime‑tuned threshold~\(\theta\):
\[
\text{out}_g = \sum_{i=0}^{W_g-1} w_i \cdot x_i, \quad
\text{prediction} = \operatorname{sign}(\text{out}_g).
\]
Weights are updated only when \(|\text{out}_g|\le\theta\) or the prediction is incorrect, ensuring rapid adaptation and bounded training costs.

Although its standalone accuracy gain is modest (\(<1\%\) BPC), the global perceptron provides a \emph{backup view} that often corrects rare, history‑spanning patterns missed by both the local model and TAGE‑SC‑L, while adding negligible footprint.

\subsection{Prediction Arbitration}
\label{sec:arbiter}

Whenever an H2P‑resident branch is fetched, \emph{all} engines fire concurrently: the baseline TAGE tables, the statistical corrector (SC), and Bullseye's two perceptrons. The final outcome is chosen by a lightweight, confidence‑based arbiter:

\begin{enumerate}[label=\roman*]
\item \textbf{Perceptron gate}: Each perceptron supplies a two‑bit \emph{conf} field that encodes (a) its running win‑rate over TAGE‑SC‑L (\textit{high} if win‑rate \(\ge\!55\%\)), and (b) whether \(|\text{output}|\!>\!\theta\) for the current instance. A perceptron asserts \textit{strong} confidence only when both tests pass.

\item \textbf{TAGE gate}: TAGE‑SC‑L asserts \textit{strong} confidence when (a) the selected TAGE component's usefulness\(\!=3\) \textbf{or} (b) the SC overrides with a magnitude \(\!>\!0\).

\item \textbf{Decision rule}: If at least one perceptron has \textit{strong} confidence and TAGE‑SC‑L does \emph{not}, Bullseye chooses the perceptron outcome. Otherwise, the arbiter defaults to TAGE‑SC‑L. This rule preserves accuracy on easy branches while allowing neural takeover only during sustained benefits.

\item \textbf{Decision rule}: If at least one perceptron has \textit{strong} confidence, Bullseye chooses the perceptron outcome. Otherwise, the arbiter defaults to TAGE‑SC‑L. This rule preserves accuracy on easy branches while allowing neural takeover only during sustained benefits.
\end{enumerate}

\subsection{Selective TAGE Filtering}
\label{sec:tage_filter}

Once a branch has delivered \(128\) consecutive correct perceptron predictions \emph{without} a single TAGE‑SC‑L win, its updates are \textbf{filtered} -- i.e., subsequent outcomes bypass all TAGE and SC tables. Filtering prevents low‑utility data from evicting well‑trained entries and cuts energy by avoiding unneeded SRAM writes. If the perceptron later falters (confidence drops below the \textit{strong} threshold), filtering is automatically revoked, ensuring that valuable geometric history is never lost permanently.  Empirically, this mechanism yields a modest accuracy gain (\(<\!0.3\%\) BPC) for negligible extra storage.

\section{Experimental Results and Analysis}

We present results for CycWpPKI and BrMisPKI metrics. Overall, Bullseye achieves an average CycWpPKI of 145.09 and an average BrMisPKI of 3.405. Table~\ref{tab:mpki_comparison} compares the average BrMisPKI of TAGE-SC-L with and without the Bullseye Predictor.

\begin{figure}[htbp]
    \centering
    \includegraphics[width=0.8\linewidth]{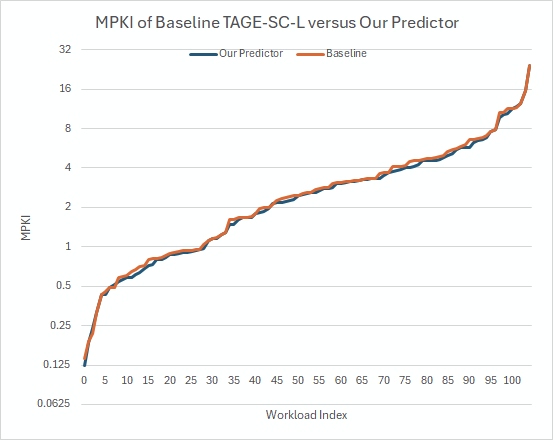}
    \caption{The BrMisPKI across all workloads for Bullseye.}
    \Description{The BrMisPKI across all workloads for Bullseye.}
    \label{fig:cyc}
\end{figure}

\begin{table}[!ht]
\centering
\begin{tabular}{|l|c|}
\hline
\textbf{Branch Predictor} & \textbf{BrMisPKI} \\
\hline
159 kB TAGE-SC-L + Bullseye (Total 187 kB)  & 3.4045\\
192 kB TAGE-SC-L & 3.4277 \\
159 kB TAGE-SC-L & 3.4513 \\
\hline
\end{tabular}
\caption{Comparison of BrMisPKI Across Predictors}
\label{tab:mpki_comparison}
\vspace{-0.2in}
\end{table}

\section{Discussion}
Bullseye identifies pathological control flow by dynamically adjusting HIT thresholds based on the current H2P population rather than absolute miss rates, leading to significant misprediction reductions. Its workload-agnostic design allows the underlying TAGE-SC-L predictor to handle typical branches, while the perceptron tier activates only for statistically qualifying branches, facilitating seamless deployment across diverse workloads without retuning. Additionally, Bullseye's lightweight H2P identification mechanism can augment other predictors. However, it does not capture data-dependent correlations, a limitation we explored but found offered only marginal accuracy gains within our area constraints, highlighting an opportunity for future H2P engines that integrate branch history with lightweight data value predictors.

\section{Conclusions}

Bullseye shows that selectively augmenting a compact 159 KB TAGE-SC-L with a lightweight, H2P-aware neural tier yields outsized accuracy gains. The H2P Identification Table (HIT) isolates the small set of static branches that dominate the misprediction tail and admits them to local- and global-history perceptrons for a gated trial. It then applies adaptive thresholds to prevent resource thrashing. A confidence-based arbiter allows neural predictions to override TAGE-SC-L only when they provide consistent benefit, while update filtering protects TAGE-SC-L from low-value training noise. Across workloads, Bullseye achieves an MPKI of 3.405.

\bibliographystyle{ACM-Reference-Format}
\bibliography{main}

\appendix

\section{Cost Analysis}

Table \ref{tab:mem_breakdown} provides a breakdown of the memory usage by each component. Between prediction time and update time the following is stored: for the local history perceptron, the local history; for the global history perceptron, the global history; and for TAGE-SC-L the requirements are unchanged from the base version with the same local and global history required to update.

\begin{table*}[!b]
\centering
\begin{tabular}{|m{0.25\columnwidth}|m{1.55\columnwidth}|m{0.10\columnwidth}|} 
\hline
 \textbf{Component} & \textbf{Details of each field of each entry and memory breakdown} & \textbf{Cost} \\
 \hline\hline
 TAGE-SC-L & Calculated using the built in TAGE-SC-L memory usage calculator & 159.34 kB \\ \hline

H2P Identification Table (HIT Cache)&
PC Tag: 10 bits (16 bit tag, but 10 bits stored with set-association) * 2\textsuperscript{6} sets * 8 ways = 5120 bits\newline 
Correct Prediction Counters: 16 bits * 2\textsuperscript{6} sets * 8 ways = 8192 bits\newline 
Incorrect Prediction Counters: 12 bits * 2\textsuperscript{6} sets * 8 ways = 6144 bits&
2.375 kB
\\ \hline
 
Global History Perceptron and FIFO&
H2P PC Tag: 62 bit PC * 16 H2P entries = 992 bits\newline
PC Queue: 62 PC bits * 64 queue entries = 3968 bits\newline 
Global History: 128 bits\newline
Weights: 12 bit precision * 128 tables * 16 H2P entries = 24576 bits\newline
Perceptron Bias: 10 bit precision * 2\textsuperscript{4} table entries * 16 H2P entries = 2560\newline
Update Thresh. Counters: (14+7) * 16 H2P entries = 336 bits \newline
Branch Management Counters: (6+8+9+16) * 16 H2P entries = 624 bits&
4.05 kB
 \\ \hline 

Local History Perceptron and FIFO&
H2P PC Tag: 62 bit PC * 32 H2P entries = 1984 bits\newline
PC Queue: 62 PC bits * 64 queue entries = 3968 bits\newline
Weights: 10 bit precision * 64 tables * table size 2\textsuperscript{8} = 163840 bits\newline
Local History: 124 bit history * 32 H2P entries = 3968 bits\newline
Perceptron Bias: 12 bit precision * 2\textsuperscript{1} table entries * 32 H2P entries = 768 bits\newline
Update Thresh. Counters: (10+7) * 32 H2P entries = 544 bits \newline
Branch Management Counters: (6+8+9+16) * 32 H2P entries = 1248 bits&
21.52 kB
\\ \hline

\multicolumn{2}{|c|}{TOTAL}  & 187.28 kB \\ \hline
\end{tabular}

\caption{Memory Usage Breakdown}
\label{tab:mem_breakdown}

\end{table*}
\end{document}